\documentclass{aa}
\usepackage{graphicx}
\usepackage{pdfpages} 
\usepackage[varg]{txfonts}
\usepackage{natbib}
\bibpunct{(}{)}{;}{a}{}{,}
\usepackage{color}
\usepackage{epsf,amsfonts,amssymb}
\usepackage{epstopdf}

\def\spose#1{\hbox to 0pt{#1\hss}}
\def\gsim{\mathrel{\spose{\lower 3pt\hbox{$\mathchar"218$}}
          \raise 2.0pt\hbox{$\mathchar"13E$}}}
\def\lsim{\mathrel{\spose{\lower 3pt\hbox{$\mathchar"218$}}
          \raise 2.0pt\hbox{$\mathchar"13C$}}}
          
\usepackage{verbatim,amscd, graphicx, amscd, mathrsfs}
\usepackage{amsmath}
\usepackage[english]{babel}
\usepackage[T1]{fontenc}
\usepackage{color}
\usepackage{listings}
\usepackage{lmodern}
\usepackage[colorlinks=true, allcolors=blue]{hyperref}
\usepackage{caption}
\usepackage{geometry}
\usepackage{graphicx}
\usepackage{multicol}
\usepackage{multirow}
\usepackage{float}
\usepackage{braket}
\usepackage{ulem}
\usepackage[utf8]{inputenc}

\newcommand{\Msol}{M_\odot}

\newcommand{\degree}{{\rm deg}}
\newcommand{\new}[1]{}
\newcommand{\LEt}[1]{}

\begin{document}

\title{Prospects for kilonova signals \\in the gravitational-wave era}

\author{R. Mochkovitch\thanks{Corresponding author: mochko@iap.fr}\textsuperscript{1}, F. Daigne\textsuperscript{1}, R. Duque\textsuperscript{1} \& H. Zitouni\textsuperscript{2}}

\authorrunning{R. Mochkovitch, et al.}
\titlerunning{Kilonovae in the gravitational-wave era}

\institute{\textsuperscript{1}Sorbonne Université, CNRS, UMR 7095, Institut d’Astrophysique de Paris, 98 bis boulevard Arago, 75014 Paris, France\\ \textsuperscript{2}PTEA Laboratory, Faculty of Science, Dr. Yahia Fares University, Médéa, Algeria}

\abstract{\LEt{ General notes: A.) You show a preference for US English grammar and spelling conventions, and I have edited accordingly throughout. B.) A\&A uses the past tense to describe specific methods used in a paper and the present tense to describe general methods as well as findings, including the findings of recent papers (within the past ten or so years). Please make sure my edits are accurate in this respect throughout the paper. See Sect. 6 of the language guide https://www.aanda.org/for-authors/language-editing/6-verb-tenses.}The binary neutron star merger gravitational-wave signal GW170817 was followed by three electromagnetic counterparts, including a kilonova arising from the radioactivity of freshly synthesized $r$-process elements in ejecta from the merger. Finding kilonovae after gravitational-wave triggers is crucial for (i) the search for further counterparts, such as the afterglow, (ii) probing the diversity of kilonovae and their dependence on the system's inclination angle, and (iii) building a sample for multi-messenger cosmology. During the third observing run of the gravitational-wave interferometer network, no kilonova counterpart was found. We aim to predict the expected population of detectable kilonova signals for the upcoming O4 and O5 observing runs of the LIGO-Virgo-KAGRA instruments. Using a simplified criterion for gravitational-wave detection and a simple GW170817-calibrated model for the kilonova peak magnitude, we determine the rate of kilonovae in reach of follow-up campaigns and their distributions in magnitude for various bands. We briefly consider the case of GW190425, the only binary neutron star merger confirmed since GW170817, and obtain constraints on its inclination angle from the non-detection of its kilonova, assuming the source was below the follow-up thresholds. We also show that non-gravitational-wave-triggered kilonovae can be a numerous class of sources in future surveys and briefly discuss associations with short bright gamma-ray bursts. We finally discuss the detection of the jetted outflow afterglow in addition to the kilonova.}


\maketitle

\section{Introduction}
The first detection of electromagnetic counterparts to a gravitational-wave (GW) event was a truly historic event \citep{Multi}. The coalescence of two neutron stars detected by the LIGO-Virgo \LEt{ Consider defining "LIGO".}instruments on August 17, 2017 (GW170817; \citealt{Abbott_GW}) was followed 1.7~s later by a weak
short gamma-ray burst (GRB) observed by \textit{Fermi} and \textit{Integral} \citep{Fermi,Integral,Abbott_GRB}. A search of the error box with optical telescopes led, after 11 hours, to the discovery of a kilonova in the spheroidal galaxy NGC 4993 at $\sim$40~Mpc \citep[][and references therein]{Multi}. Then, after respectively 9 and 16 days the afterglow was detected in X-rays with \textit{Chandra} \citep{Troja} and in radio with the Very Large Array \citep[VLA;][]{Hallinan}. The afterglow light curve was atypical, with a steady rise to a maximum at about 170 days post-merger \citep{RiseX,DAvanzo,Fading,Resmi,Mooley}. While such a behavior could result from either a radial or an angular structure of the ejecta  \citep[e.g.,][]{Gill}, very long baseline interferometry observations showing a displacement of the unresolved source by about 2.5 mas in 5 months \citep{Mooley_VLBI,Ghirlanda} provided firm evidence for the latter. Joint fits to the afterglow photometry and imagery show that GW170817 was observed under an inclination angle of $15^{+2.5}_{-1.7}\,\degree$ \citep{Mooley,Ghirlanda}\footnote{Throughout this paper we systematically state median 90\% credible intervals. When primary sources state results with other credible intervals, we scaled such intervals assuming a Gaussian distribution.}.

The kilonova dubbed AT~2017gfo reached a peak magnitude of $\sim\,17$ in the $g$, $r$, $i$, and $z$ bands and was followed for two to three weeks in the infrared ($z$ to $K$ bands), where the decline is shallower than in the visible. 
The data are well fit by the combination of a 
blue, mostly polar component, which declines faster, and a more isotropic red component.
The red component results from the high opacity lanthanide-rich material tidally ejected during coalescence or blown off from the accretion disk
\citep{Cowperthwaite,Tanvir,Villar,Tanaka}. Different possible origins have been considered for the blue, lanthanide-poor component, which is expected to be present only if the central core does not directly
collapse into a black hole \citep{Metzger}
\footnote{In AT~2017gfo, a third ``purple component'' with an intermediate lanthanide fraction can be added to improve the fit \citep{Cowperthwaite,Villar}; however, a thorough preference for a three-component model has not been established so far.}.
In most cases, the core is probably short lived and collapses into a black hole. In the case of GW170817, the eventual GW signal of the black hole ring-down was too weak to confirm
if or when this collapse had occurred \citep{Abbott_GW}. 
This single event represented a breakthrough in the understanding of merger physics and the study of the origin of heavy elements in the Universe \citep[for a review, see][]{10.3389/fspas.2020.00027}. Also, it allowed the first standard siren measurement of the Hubble constant \citep{AAAAA+2017,2019NatAs...3..940H}.

During the third LIGO-Virgo-KAGRA \LEt{ Consider defining.}Collaboration (LVKC) observing run, O3, the only confirmed binary neutron star merger so far (i.e., GW190425) \LEt{ We do not allow the use of "e.g." or "i.e." within the main text (in parentheses or within figure/table captions is fine).} was located at $159^{+69}_{-71}$ \,Mpc \citep{GCN,LVC_GW190425}. 
With only one of the LIGO interferometers taking data at the time of the event, the GW sky map was very large, nearly $7500\,\degree^2$. It could only be partially explored by the various follow-up efforts. The Pan-STARRS and Zwicky Transient Facility (ZTF) telescopes achieved the largest coverage of the event \citep{PS2}, and no kilonova was found. It remains unknown if the kilonova was weaker than the detection limits or simply outside the area covered by the searches.

Following the premature end of O3 at the end of March 2020, the GW detectors are expected to resume operations in mid-2022 with a binary neutron star merger range increased by about 50\% \citep{LRR}. The GW discovery rate of binary neutron star mergers should then reach $10^{+52}_{-10}$ per calendar year, a factor of ten larger than during O3. The participation of KAGRA in this run will reduce the average sky surface and volume of searches for kilonova counterparts. However, these kilonovae will be located at larger distances on average, possibly impeding their detection. 

The goal of this paper is to obtain the expected distributions of various physical parameters for kilonovae that should be detected in association with GW signals of merging binary neutron stars during O4 and beyond. Using a simple parametrization of the kilonova magnitude as a function of viewing angle in different spectral bands, we obtain the distributions (i) in magnitude for visible and near-infrared bands and (ii) in viewing angle for different limiting magnitudes of the kilonova follow-up search. We also estimate the corresponding discovery rate, again for different limiting magnitudes.

In the case of GW190425, we show how a constraint on the viewing angle can be obtained from the lack of counterparts in three bands. We derive this constraint by assuming that the source was indeed located in the areas searched during follow-up efforts, but below detection threshold. \LEt{ Single-sentence paragraphs should be avoided. Consider combining this paragraph with the proceeding or preceding paragraphs.}

When a kilonova is found, the sky location is known with an arcsecond accuracy, which allows searching for the afterglow in X-rays or radio. 
We calculate the fraction of sources that can be detected in radio in addition to the kilonova with the VLA as well as their distribution in viewing angle.
In the visible, the afterglow is likely to be initially outshone by the kilonova. This is expected as long as the viewing angle is larger than the opening angle of the jet core, which is 
most probable when the alert is given by the GW detectors.

This paper is organized as follows: In Sect.~\ref{sec:2} we first obtain the distributions in distance and viewing angle of the neutron star merging events detected in GWs. Our simplified parametrization of the kilonova magnitude as a function of viewing angle is presented in Sect.~\ref{sec:knmagangle}. The resulting distributions in magnitude and viewing angle of the detectable kilonovae are shown in Sect.~\ref{sec:knpop} together with the constraints that can be obtained on the viewing angle for GW190425. The possibility to observe the afterglow when the kilonova has been found is discussed in Sect.~\ref{sec:radioaft}. Finally, our results are discussed
in Sect.~\ref{sec:discussion}, and Sect.~\ref{sec:conclusion} is the conclusion.

\section{Kilonova distance and viewing angle distributions}
\label{sec:2}
\subsection{Distribution in distance}
\begin{figure*}
\resizebox{\hsize}{!}{\includegraphics{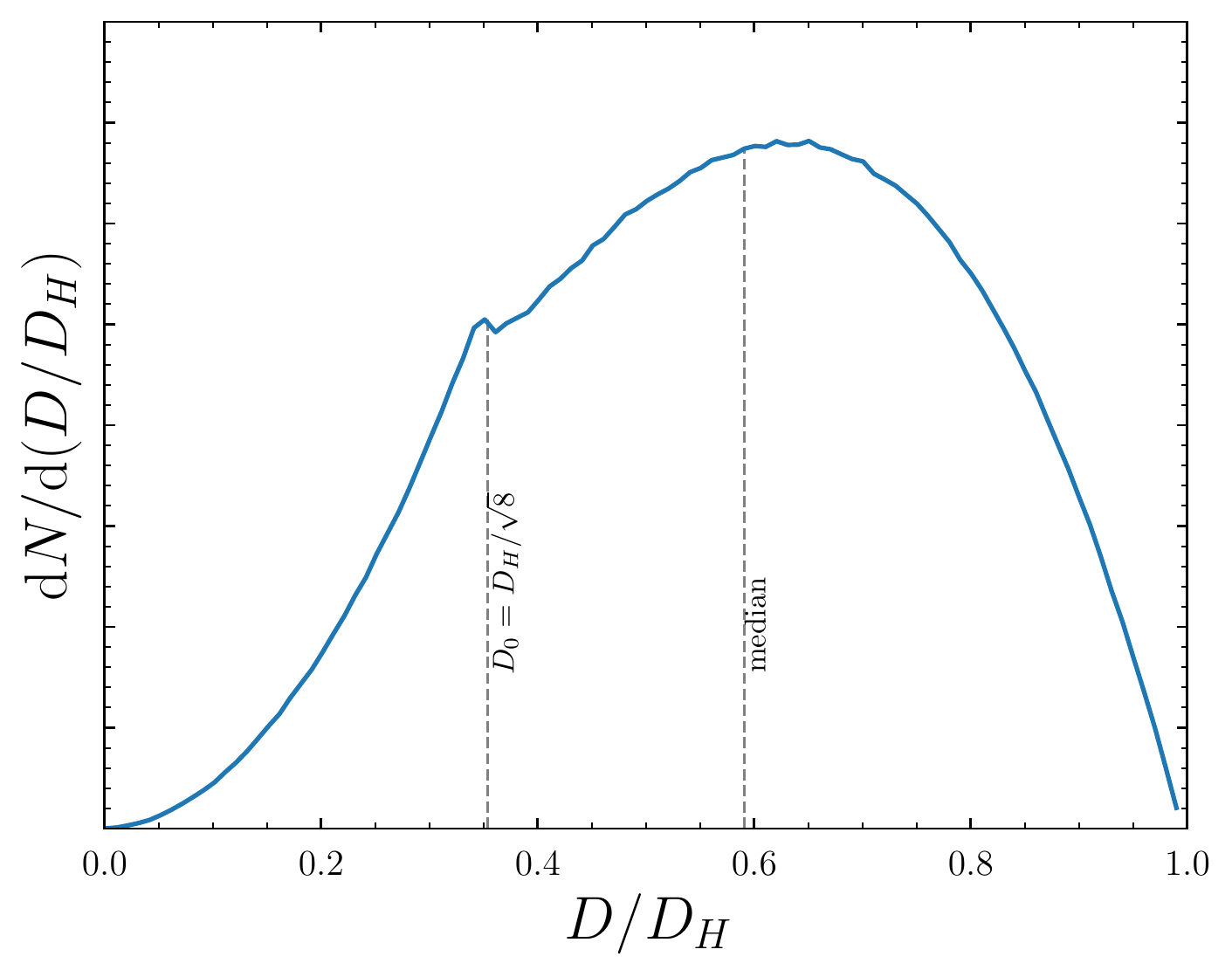}{\includegraphics{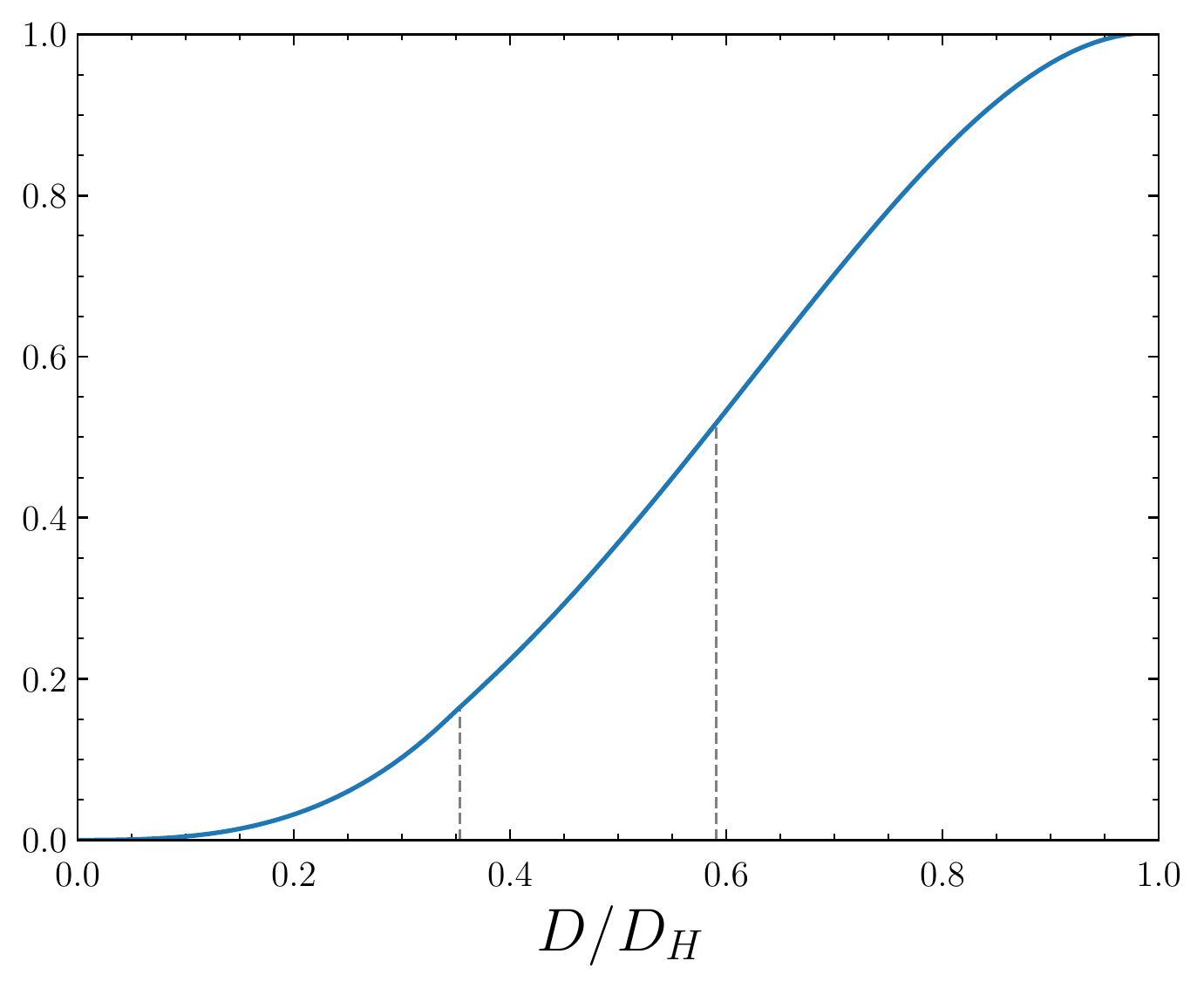}}}
\resizebox{\hsize}{!}{\includegraphics{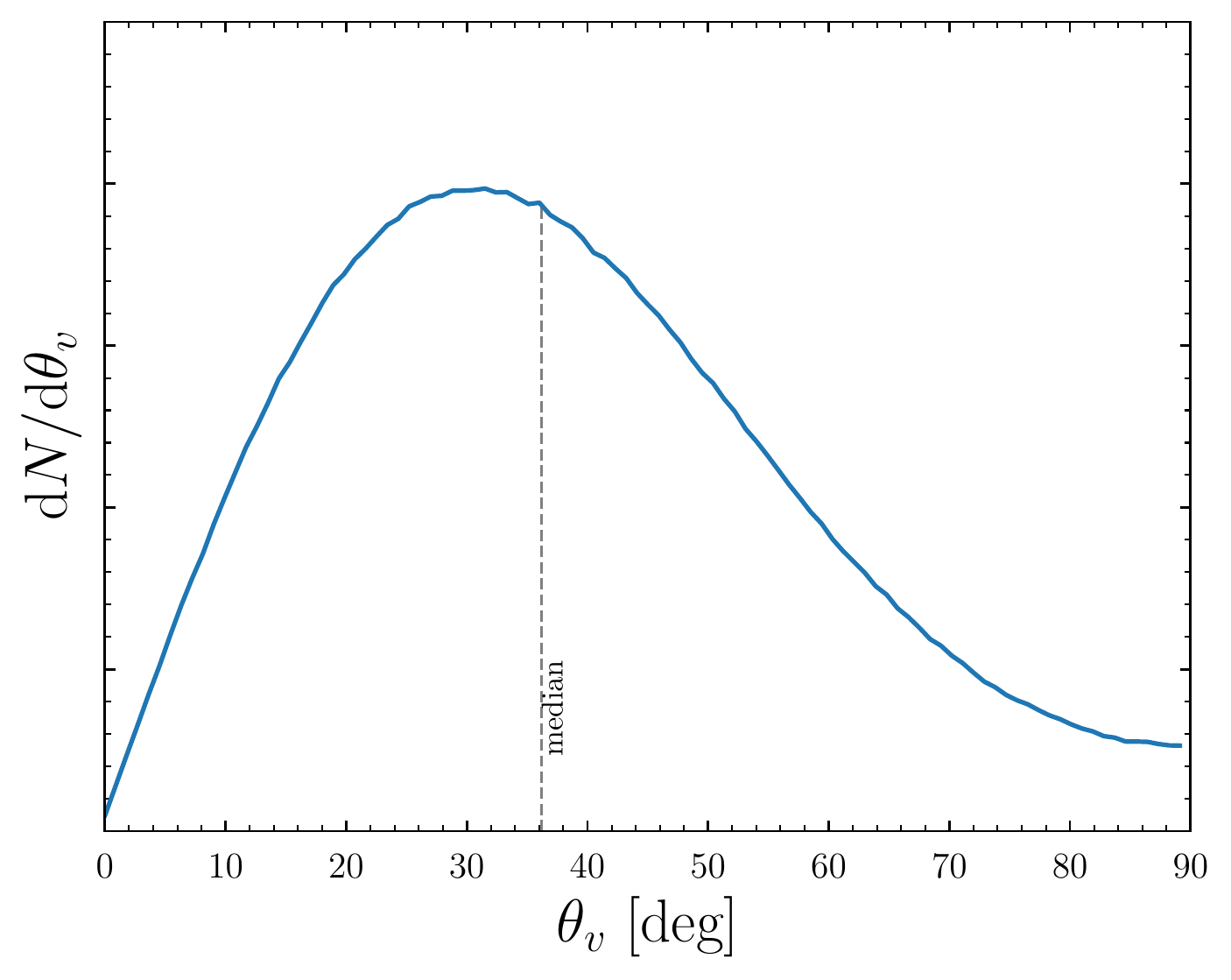}{\includegraphics{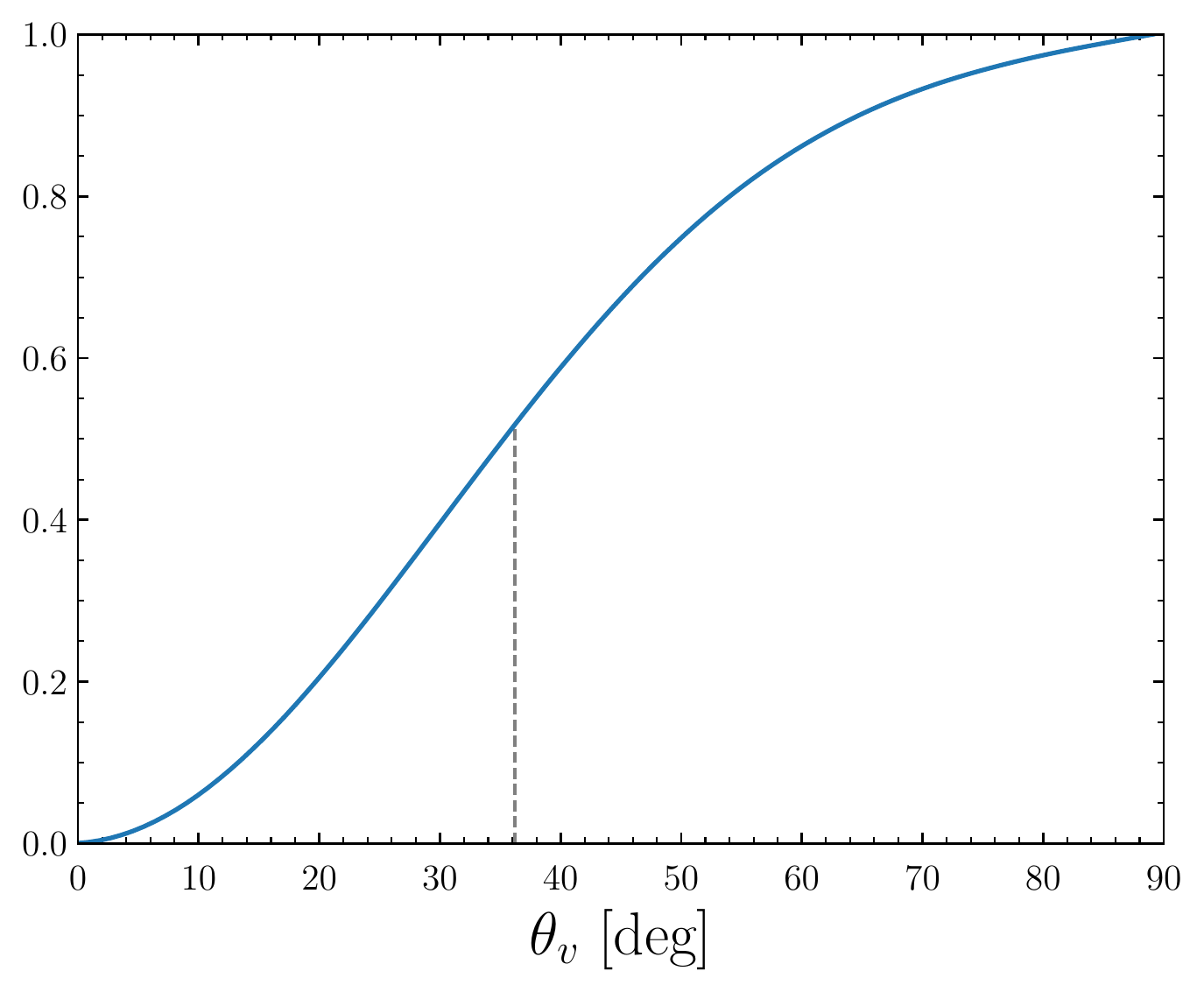}}}
\caption{Differential (left) and cumulative (right) distributions in distance (top) and viewing angle (bottom) of the GW triggers. We also indicate the median values and $D_0$, the distance under which no selection in viewing angle occurs. The cusp at $D_0$ in the differential distribution in distance is nonphysical and a consequence of the simplified nature of the adopted GW detection criterion. 
}
\label{fig:GW}
\end{figure*}

Since we are interested in kilonovae in association with a GW signal, their distance and viewing angle distributions simply follow those of GW-triggered neutron star merger events. For a given sky-position-averaged horizon $D_H$, corresponding to a binary neutron star having its rotation axis pointing toward the observer, detection at viewing angle $\theta_{v}$ is possible only at distances $D$ such that \LEt{ such that?} \citep{Schutz}
\begin{equation}
\frac{D}{D_H} \le \sqrt{\frac{1 + 6\cos^2 \theta_{v} + \cos^4 \theta_{v}}{8}}\, .
\label{eq:gw}
\end{equation}
Then, for $D\le D_0=D_H/\sqrt{8}$,
all sources are detected, while for $D_0<D\le D_H$, they are progressively lost until,
for $D_H$, only those pointing directly at the observer remain. The resulting distribution in distance is represented in Fig.~\ref{fig:GW}a, which shows a maximum at $D/D_H=0.63$. Figure~\ref{fig:GW}b gives the corresponding cumulative distribution.

In Table~\ref{tab:KNparam} we state our assumed sky-position-averaged horizons for past and upcoming GW observing runs.
These were taken from \cite{LRR} and correspond to 1.4+1.4 $\Msol$ binary neutron star systems. For the particularly massive GW190425, we adapted the horizon value (see Sect.~\ref{sec:4.4} for details).

\begin{table}[!h]
\begin{center}
\resizebox{\linewidth}{!}{\begin{minipage}{9.75cm}
\begin{center}
\begin{tabular}{|lr|lrr|}
\hline
Run & $D_H$  [Mpc] & Band & $M_{\lambda,0}$ & $\Delta M_{\lambda}$   \\
\hline
 O3 & 157 & $g$  & $-16.3$ & $7$  \\
O4 & 229 & $r$  & $-16.3$ & $4$ \\
O5 & 472 & $i$ & $-16.4$ & $3.5$  \\
O3@GW190425 & 181 & $z$  & $-16.5$ & $2.5$ \\
\hline
\end{tabular}
\end{center}
\end{minipage}}
\end{center}
\caption{
Horizon distances assumed for the various GW observing runs, as used in the detection criterion in Eq.~\ref{eq:gw}, and 
parameters for the kilonova peak absolute AB magnitude dependence on the viewing angle, as given in Eq.~\ref{eq:KNmag}.
}
\label{tab:KNparam}
\end{table}

\subsection{Distribution in viewing angle}
The distribution in viewing angle of the gravitationally detected sources is represented in Figs.~\ref{fig:GW}c and~\ref{fig:GW}d. It peaks at $\theta_{v}\sim 30\,{\degree}$, with an average value $\langle\theta_{v}\rangle=38\,\degree$.
The fraction of sources in the conservative interval $10\,\degree<\theta_{v}<20\,\degree$ corresponding to GW170817 is 14\% (increased to 27\% for $5\,\degree<\theta_{v}<25\,\degree$).

Had GW170817 occurred during the O3 run, with $D_H = 157$\,Mpc, the distance to the source would have verified that $D_{\rm 170817} \sim 40\,{\rm Mpc}< D_0$. Therefore, any merger at this distance would have been detectable, regardless of the inclination angle. In this case, the expected rate of binary neutron star mergers up to the distance of GW170817
is simply given by
\begin{equation}
R=\tau_\mathrm{BNS}\times {4\pi\over 3} {D_{\rm 170817}}^3\, ,
\end{equation}
where $\tau_\mathrm{BNS} = 320^{+490}_{-240} $ Gpc$^{-3}$yr$^{-1}$ is the local binary neutron star merger rate \citep{2020arXiv201014527A}. This leads to an average rate of one event every $R^{-1} = 12_{-7}^{+36}\,{\rm yr}$. 

GW170817 not only was a nearby event but had a low inclination angle,
$\theta_v^{\rm 170817} < \theta_v^{\rm 170817, max} \sim 18\,\degree$,
according to the very long baseline interferometry observations \citep{Mooley_VLBI,Ghirlanda}.
The detection of the radio afterglow and source proper motion was possible only up to a viewing angle of $\theta_v^{\rm AG, max} \sim 40\,\degree$ \citep[e.g.,][]{Duque}. Requiring $\theta_v\le \theta_{v}^\mathrm{AG, max}$ -- to get a rich multi-messenger data set with an inspiral signal as well as kilonova and afterglow photometry and imagery data  --
therefore leads to a rate of approximately\LEt{ Verify that your intended meaning has not been changed.}
\begin{equation}
R'=R\times (1-{\rm cos}\,\theta_v^{\rm AG, max})\, ,
\end{equation}
that is, an average rate of one 
event every 
${R'}^{-1} = 50_{-31}^{+149}\,{\rm yr}$.

The detection of the short GRB may require a even smaller viewing angle, $\theta_\mathrm{v} \le \theta_v^{\rm GRB, max}$ with $\theta_v^{\rm GRB, max}\simeq \theta_v^{\rm 170817, max}$, as GRB170817A was detected only at the $\sim 5\sigma$ level by the Gamma-Ray Burst Monitor aboard \textit{Fermi} \citep{Fermi}. 
Requiring $\theta_\mathrm{v} \le \theta_v^{\rm GRB, max}$ to get a full GW170817-like multi-messenger data set, including the short GRB, leads to an even lower rate,
$R''=R\times (1-{\rm cos}\,\theta_v^{\rm 170817, max})$, that is, one 
event every ${R''}^{-1} = 239_{-146}^{+713}\,{\rm yr}$.

During O2 ($D_H \sim 86\, \mathrm{Mpc}$; \citealt{LRR}), when $D_{\rm 170817}$ was not smaller than $D_0 \sim 30~{\rm Mpc}$, these rates were even lower. These numbers illustrate how lucky we were to detect GW170817 so early and how long we may have to wait to observe another equivalent event.

\begin{table*}
\begin{center}
\begin{tabular}{|l|lll|lll|lll|}
\hline
AB Mag. range& \multicolumn{3}{c}{$<18$} & \multicolumn{3}{c}{$18-20$} & \multicolumn{3}{c}{$>20$} \vline\\
Band & O3 & O4 &  O5 & O3 & O4 & O5 & O3 & O4 & O5 \\
\hline
$g$ &   2.5 &  0.81 & $< 0.1$ &    24 &    11 &   1.4 &    74 &    88 &    99 \\
$r$ &   4.1 &   1.3 &  0.15 &    39 &    19 &   2.3 &    57 &    80 &    98 \\
$i$ &   5.6 &   1.8 &  0.21 &    48 &    25 &   3.1 &    46 &    73 &    97 \\
$z$ &   8.9 &   2.9 &  0.33 &    65 &    38 &   4.9 &    26 &    59 &    95 \\
\hline
\end{tabular}
\end{center}
\caption{Percentage fraction of kilonovae associated with GW triggers in the three magnitude intervals $m<18$, $18<m<20$, and $m>20$. Figures correspond to observing runs O3, O4, and O5.}
\label{tab:KNfrac}
\end{table*}

\section{Kilonova magnitude  dependence to viewing angle}
\label{sec:knmagangle}

The kilonova magnitude at the peak depends on the distributions of the mass, velocity, and composition of the ejected material as well as on the viewing conditions: distance and viewing angle. The ejection is anisotropic with neutron-rich, dynamical ejecta in the equatorial plane,
where the formation of lanthanides leads to a large opacity while a relatively neutron-poor wind of lower opacity is blown in the polar direction \citep{2016ARNPS..66...23F,Metzger,10.3389/fphy.2020.00355}. This wind is expected to be present when a short-lived massive neutron star is formed before collapsing to a black hole, but probably not in the case of a direct collapse.
The lanthanide-rich ejecta produces the
``red kilonova,'' which peaks in the near-infrared, while the neutron-poor wind is responsible for the ``blue kilonova'' at optical wavelengths. The blue kilonova declines on a timescale of one day, whereas the timescale is one week for the red component.

For our population model, our default scenario assumes that all kilonovae have a quasi-isotropic red component and a polar blue component. We obtain the peak absolute AB magnitude at a given wavelength and viewing angle from the following simple parametrization:

\begin{multline}
M_{\lambda,\theta_{v}}= \\
\left\{ \begin{array}{ll} M_{\lambda,0}+\Delta M_{\lambda}\left(\frac{1-\cos  \theta_{v}}{1-\cos\,\theta_0}\right)+\delta M_{\lambda}, & \hskip 0.2cm \theta_v\le \theta_0 \\
M_{\lambda,0}+\Delta M_{\lambda}+\delta M_{\lambda}, & \hskip 0.2cm\theta_0\le \theta_v, \end{array}\right.
\label{eq:KNmag}
\end{multline}
where $M_{\lambda,0}$ is the peak absolute magnitude for a polar viewer and $\Delta M_{\lambda}$ is the amplitude of the polar effect. The\LEt{ Avoid beginning a sentence with an acronym, abbreviation, number (unless written out), formula, or symbol.} $\delta M_{\lambda}$ represents the intrinsic (i.e., non-viewing-angle-related) variability in kilonova magnitudes linked to the abovementioned ejecta properties and, in turn, to the progenitor component masses and spins. 
For $\theta_0 = 60\,\degree$, we find that the linear-in-$\cos\theta_v$ form of Eq.~\ref{eq:KNmag} reproduces the trends of sophisticated kilonova modeling work (e.g., \citealt{WKFRE+2018}, \citealt{2020ApJ...889..171K}, and the ``asymmetric model'' of \citealt{VGBMC+2017}). 
We chose $\delta M_{\lambda}$ to be uniformly distributed in $[-1, 1]$, reproducing the expected variability in kilonova magnitude stemming from variability in ejecta mass, velocity, and opacity \cite[][Eq. 33]{WKFRE+2018}.   
The difference in magnitude between equatorial and polar views is moderate in the infrared but increases rapidly in the visible, already reaching about 4 mag in the $r$ band. This is mainly due to the stronger anisotropy of the blue component.
To calibrate this expression, we used AT~2017gfo, assuming $\theta_{v}= 15\,\degree$, as mentioned in the introduction. Corresponding values can be found in Table~\ref{tab:KNparam}.

Calibrating Eq.~\ref{eq:KNmag} with AT~2017gfo supposes that this transient was representative of the kilonova population. 
This is the minimal hypothesis one can make while waiting for the number of kilonovae with robust angle measurements to increase in the future. We note that AT~2017gfo could have been brighter or dimmer than the average of the population our model seeks to encapsulate. We briefly indicate below how our results might change if this is indeed the case.

The polar ejecta may not be produced in all mergers, depending for instance on the post-merger formation of a massive neutron star before its collapse into a black hole \citep[see, e.g.,][]{Metzger}. As such, we also consider below 
the possibility that a fraction of the kilonova population lacks the blue component, which would affect kilonova brightnesses in the bluer bands (see a preliminary luminosity function
in \citealt{Ascenzi} and related discussions in \citealt{Gompertz} and \citealt{Kasliwal_LF}). \LEt{ single-sentence paragraph.}

\section{Resulting kilonova population}
\label{sec:knpop}
\subsection{Apparent magnitude}
\begin{figure*}
\resizebox{\hsize}{!}{\includegraphics{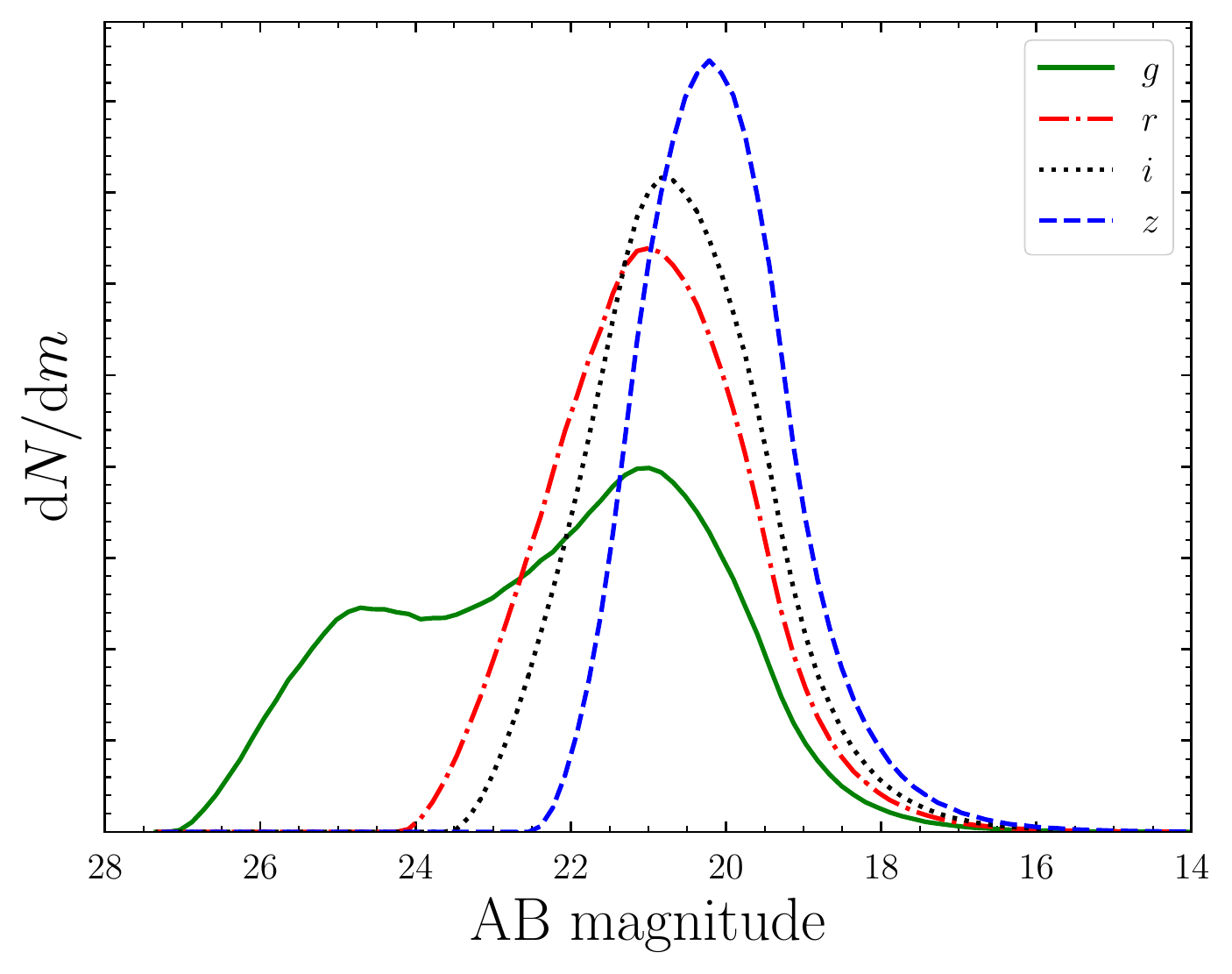}{\includegraphics{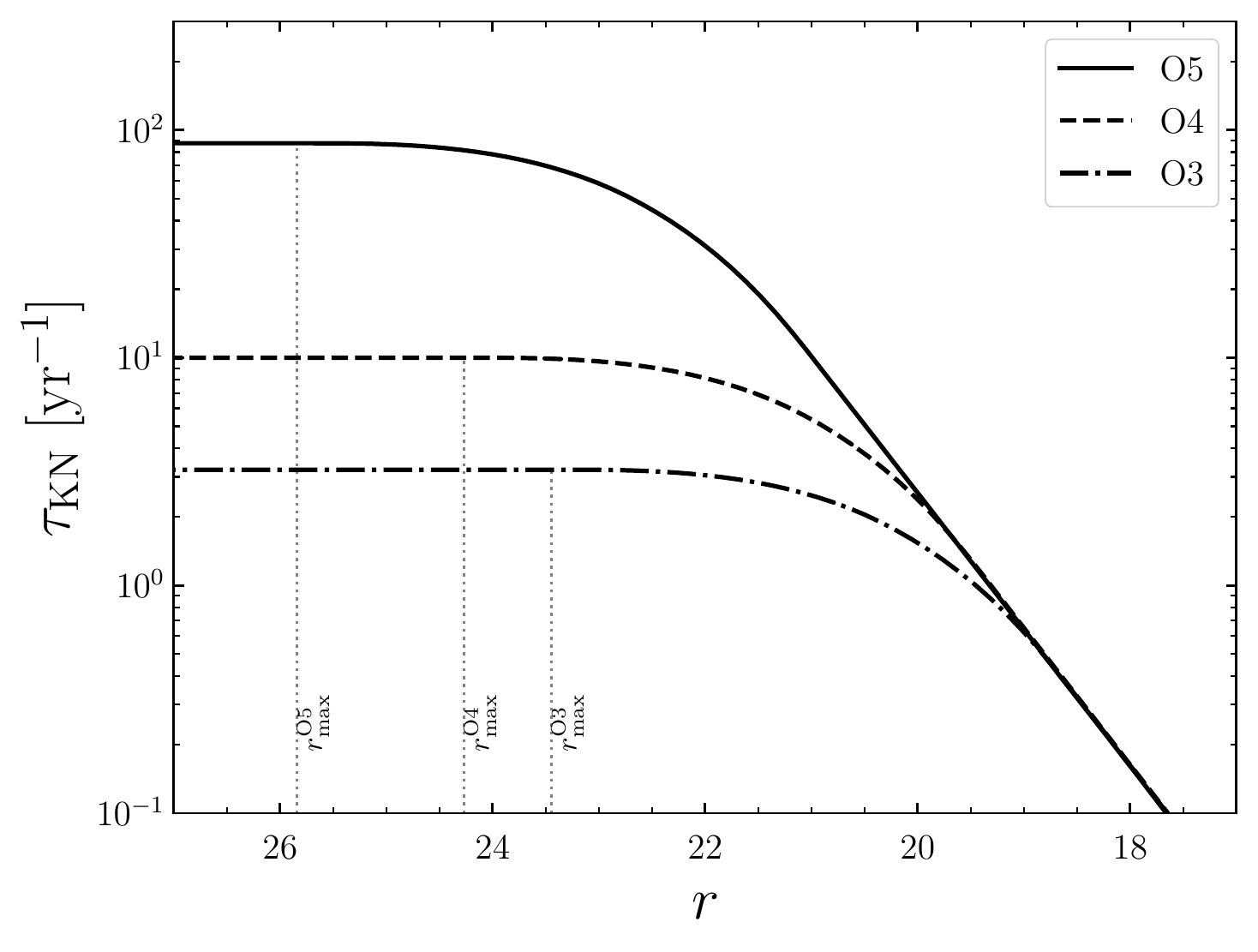}}}
\caption{Distribution of peak kilonova magnitudes and kilonova detection rate as a function of threshold. Left: Distribution of the peak AB magnitude in the $g$, $r$, $i$, and $z$ bands predicted for kilonovae associated with GW triggers during O4. Right: Rate of kilonovae brighter than a given $r$ magnitude associated with GW detections during O3, O4, and O5, 
assuming a GW neutron star coalescence detection rate of $\tau_\mathrm{BNS,GW}=10\,{\rm yr}^{-1}$ for O4 \citep{LRR}. The bright end of the distribution ($r<19$) is well fit by Eq.~\ref{eq:KNrate}.}
\label{fig:KNmag}
\end{figure*}

From the known distance and viewing angle distributions and our adopted parametrization for the magnitude 
(Eq.~\ref{eq:KNmag}), we can readily obtain the distribution of apparent AB magnitudes for kilonovae associated with GW detections. It is shown in Fig.~\ref{fig:KNmag}a for the $g$, $r$, $i$, and $z$ bands for the O4 observing run. If AT~2017gfo was in fact brighter than the average population, all the curves will have to be shifted by the corresponding difference, $\delta {\rm mag}=\langle m \rangle - m_{170817}$. Changing the GW horizon implies an interplay between the maximum detection distance for the kilonova and the GW and thus does not result in a simple shifting of the magnitude distribution. However, we found that, to a good approximation, changing from O4 to O5, the magnitude distribution is shifted by about $5\,{\rm Log}(D_{H,O5}/D_{H,O4})=1.6$ mag.


The distribution of kilonovae in different magnitude ranges is summarized in Table~\ref{tab:KNfrac} for three GW sensitivity hypotheses: O3, O4, and O5. It can be seen that there are very few kilonovae with $m<18$ in all cases beyond O3. We note that recalibrating Eq.~\ref{eq:KNmag} assuming that AT~2017gfo was one magnitude brighter than average leads to dividing the expected fractions in the $< 20$ magnitude ranges for O4 by approximately three.

The kilonova magnitudes depend on the different merger ejecta and their physical conditions. The blue kilonova component is likely linked to neutrino- or magnetohydrodynamic-viscosity-driven winds from the transient remnant product and the accretion disk around such a product \LEt{ Verify that your intended meaning has not been changed, i.e., that this is what "latter" referred to.} \cite[][and references therein]{2019ApJ...876..139G}.

It is possible that in some systems blue-enhancing ejection episodes are less effective (e.g., because of a short-lived merger remnant), leading to a lack of a blue kilonova component. We briefly consider this possibility, without seeking to know the fraction of these cases in the population.

If a fraction $f_{\rm red}$ of the kilonovae lack the blue component, a simple approximation consists in stating 
that such kilonovae will be dimmer and thus transferred from the two brightest magnitude groups to the $m>20$ group. This leads to the following for all bands:
\begin{equation}
\begin{array}{lll}
f_{<18}&\sim& f_{<18}^0\times (1-f_{\rm red})\\
f_{18-20}&\sim& f_{18-20}^0\times (1-f_{\rm red})\\
f_{>20}&\sim& f_{>20}^0\times (1-f_{\rm red})+f_{\rm red}
\end{array}
,\end{equation}
where the $f^0$ fractions are those listed in Table~\ref{tab:KNfrac}.
We tested these approximations with our kilonova population model by emulating the absence of the blue component in a fraction $f_{\rm red}$ of the synthetic kilonovae. We did this by adopting the $\theta_v > \theta_0$ case of Eq.~\ref{eq:KNmag} for all viewing angles, as if only the red component were present. As the blue component affects the $g$, $r$, and $i$ bands more, these expressions represent reasonable approximations of the exact results, while in the $z$ band they somewhat overestimate the number of events that change from the $<20$ to the $>20$ magnitude groups.

The expected rates of kilonovae brighter than a given limiting $r$ magnitude are shown in Fig.~\ref{fig:KNmag}b for O3, O4, and O5, normalized to a GW neutron star coalescence detection rate of $\tau_\mathrm{BNS,GW}=10\,{\rm yr}^{-1}$ for O4 \citep{LRR}. At the bright end of the distribution (i.e., $r<19$), a fit to Fig.~\ref{fig:KNmag}b shows that the rate approximately follows:
\begin{equation}
\log\frac{\tau_{\rm KN}}{{\rm yr}^{-1}} = 0.60 \times r_{\rm lim} - 11.6 + \log(1 - f_{\rm red})
\label{eq:KNrate}
,\end{equation}
where $r_{\rm lim}$ is the limiting magnitude in the $r$ band.

As an illustration, with $f_{\rm red}=0$ and $f_{\rm red}=0.2$ we expect one kilonova brighter than $r=19$ every 1.6 years and every 2.0 years, respectively, and one brighter than $r=18$ every 6.3 years and 7.9 years, independently of any future improvement in the sensitivity of GW detectors. On the other hand, the rate of kilonovae detectable by a follow-up with a limit magnitude $r_\mathrm{lim}=21$ is increased by a factor of $\sim 3$ between O3 and O4.

In Fig.~\ref{fig:KNmag}b we also show the maximum $r$-band magnitude of any kilonova associated with
a GW trigger for O3, O4, and O5, denoted by $r_{\rm max}$. These magnitudes are the search depths required to recover 100\% of the kilonovae. Because our peak magnitude dependence with viewing angle saturates at $\theta_0 = 60\,\degree$, these maximum-magnitude events have $\theta_v = \theta_0$ and are placed at the largest distance to which the GW signal can be detected at this angle (i.e., at $D / D_H = \sqrt{(1 + 6 \cos^2 \theta_0 + \cos^4 \theta_0)/8} \sim 0.55$).

\subsection{Distribution in viewing angle for different limiting magnitudes}

\begin{figure}
\centering
{\includegraphics[width=7.5cm]{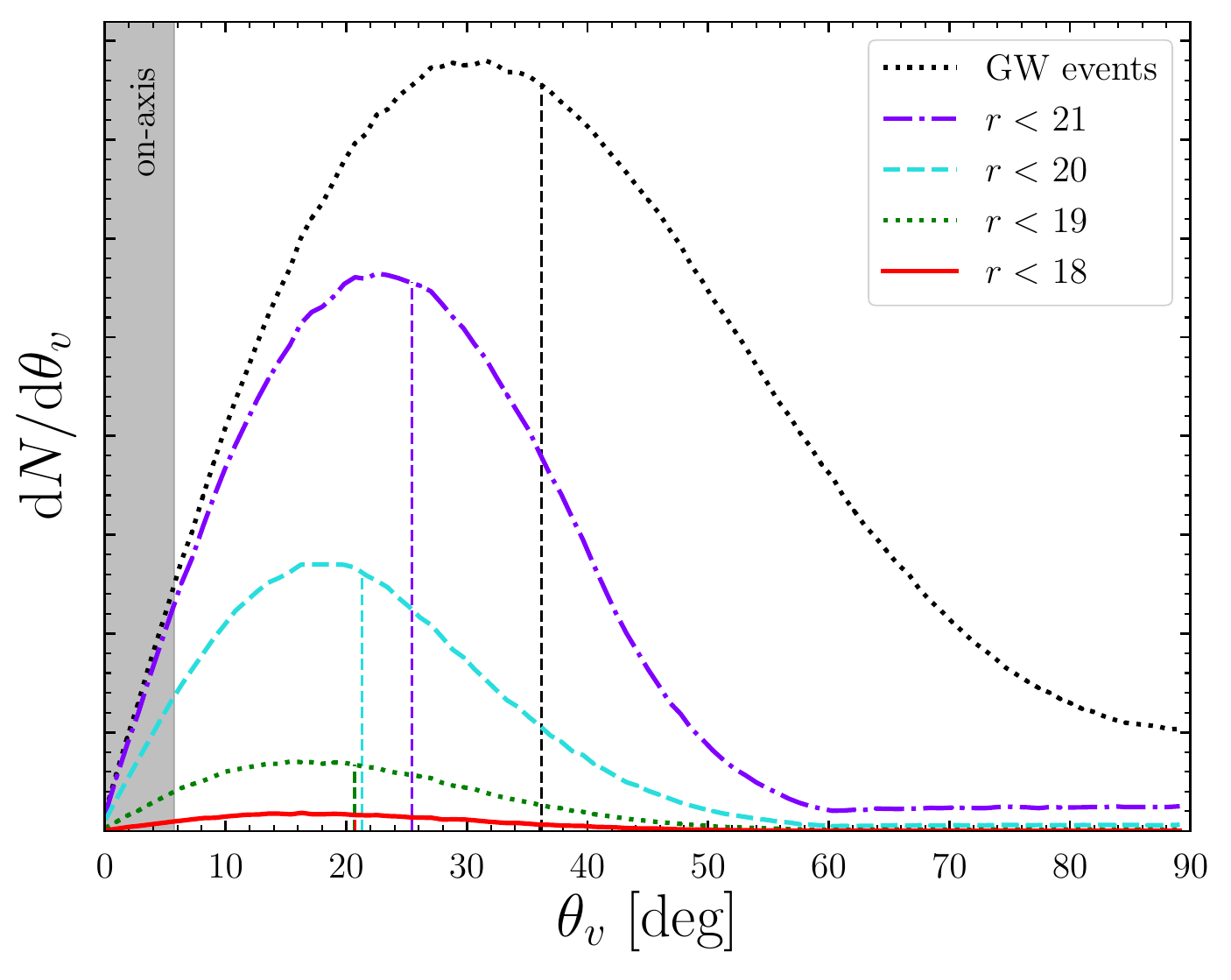}}
\caption{Distribution in viewing angle predicted for detectable kilonovae associated with GW detections during O4 and for limiting $r$-band magnitudes 21, 20, 19, and 18. The vertical dashed lines represent the median.}
\label{fig:KNorientation}
\end{figure}

The distribution in viewing angle of the kilonovae associated with binary neutron star merger triggers that are brighter than a given limiting magnitude is shown in Fig.~\ref{fig:KNorientation} for O4. As the limiting magnitude decreases, the median kilonova viewing angle -- close to $36\,\degree$ in the entire population of GW triggers -- significantly decreases: $26\,\degree$ for an $r$-band limiting magnitude of $21$ and $21\,\degree$ for all $r_{\rm lim}$ smaller than $20$.

\begin{figure*}
\resizebox{\hsize}{!}{\includegraphics{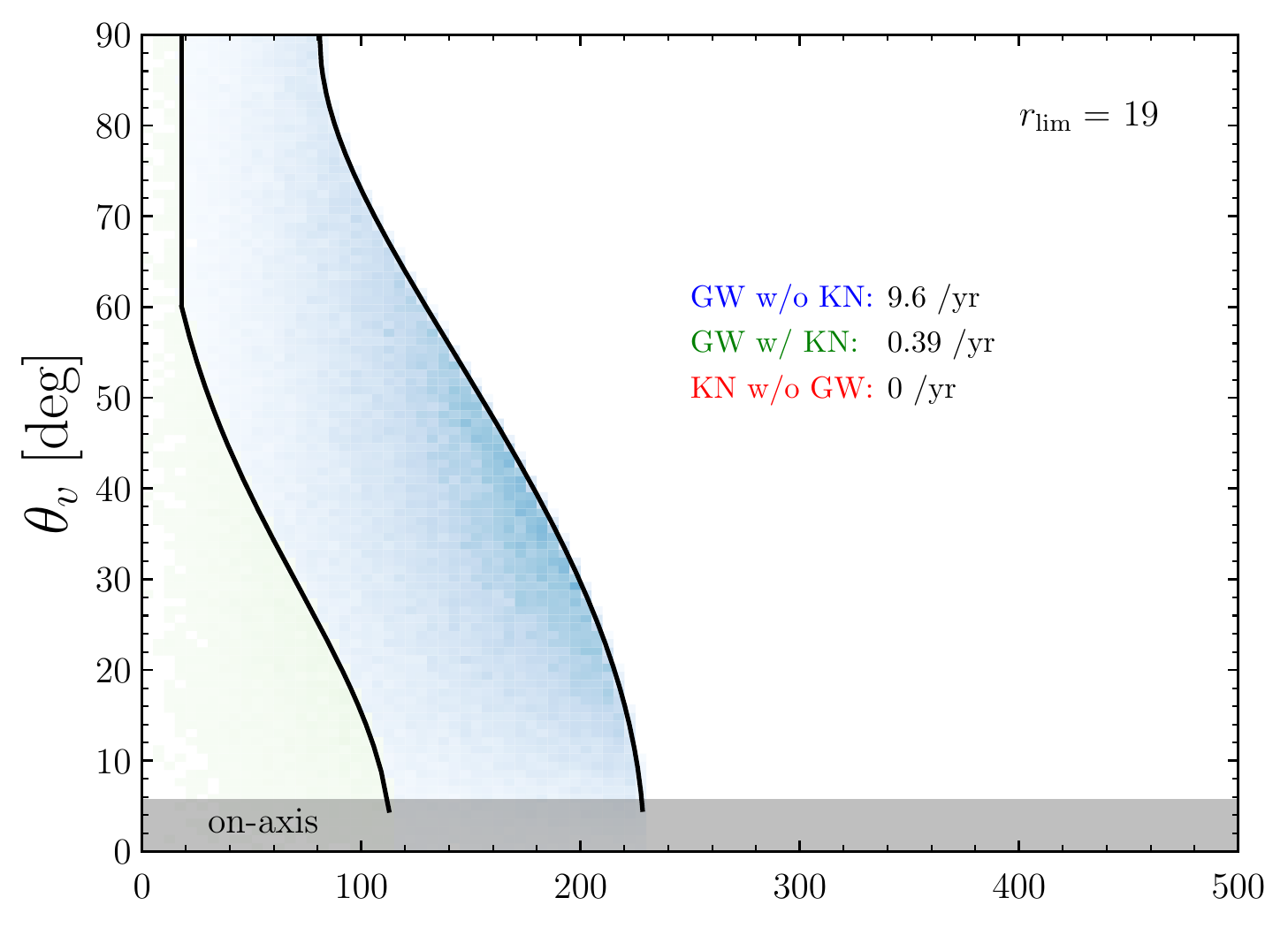}{\includegraphics{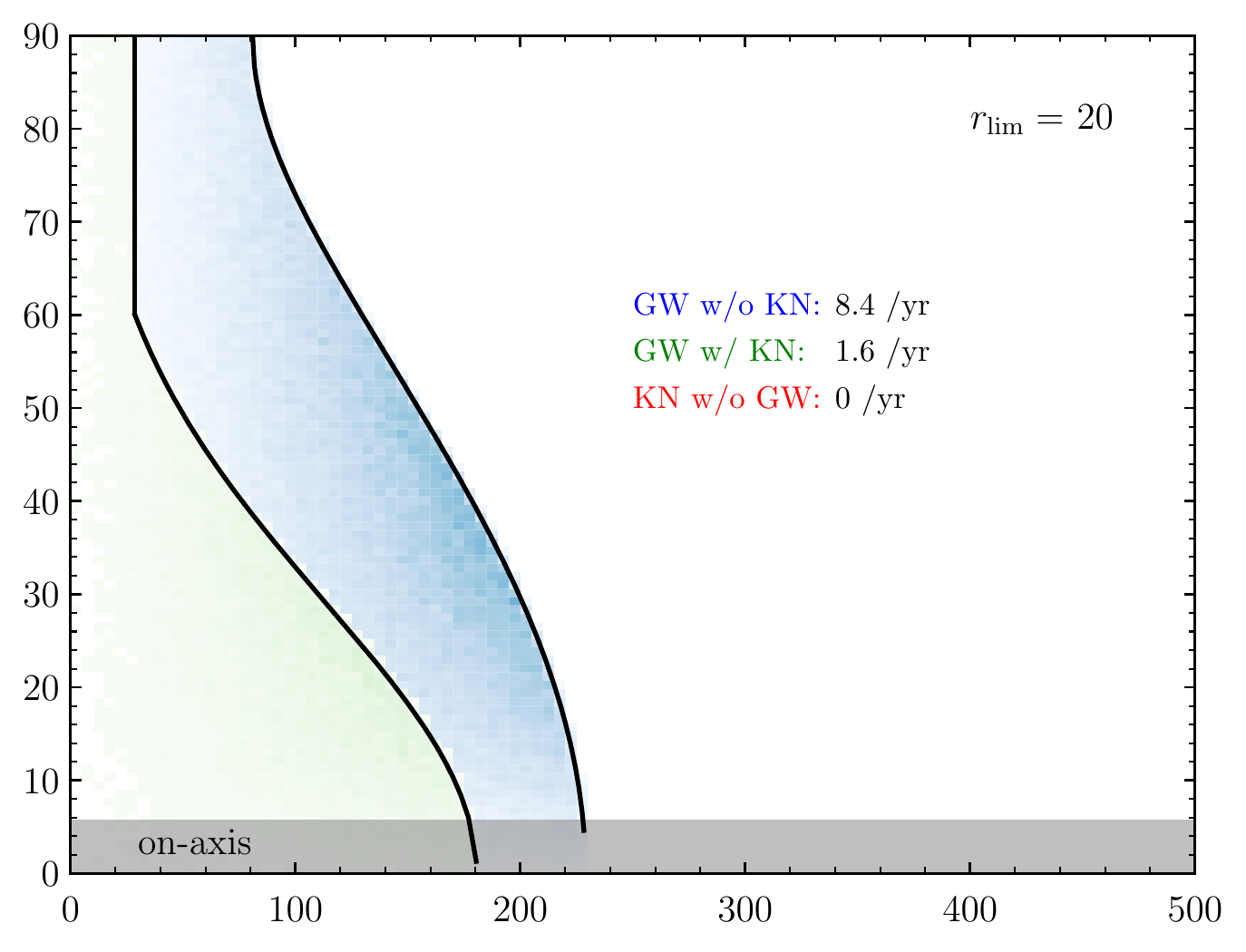}}}
\resizebox{\hsize}{!}{\includegraphics{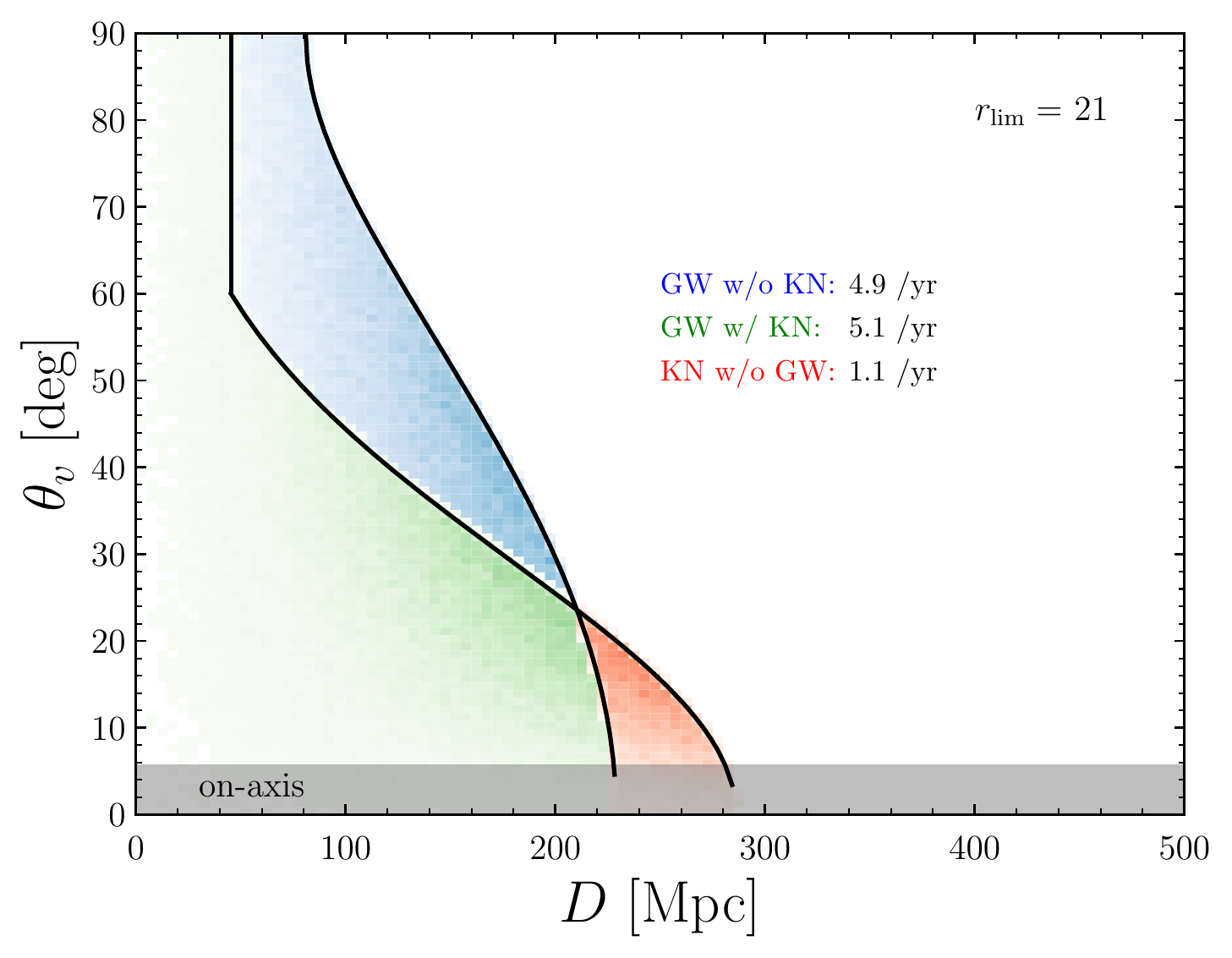}{\includegraphics{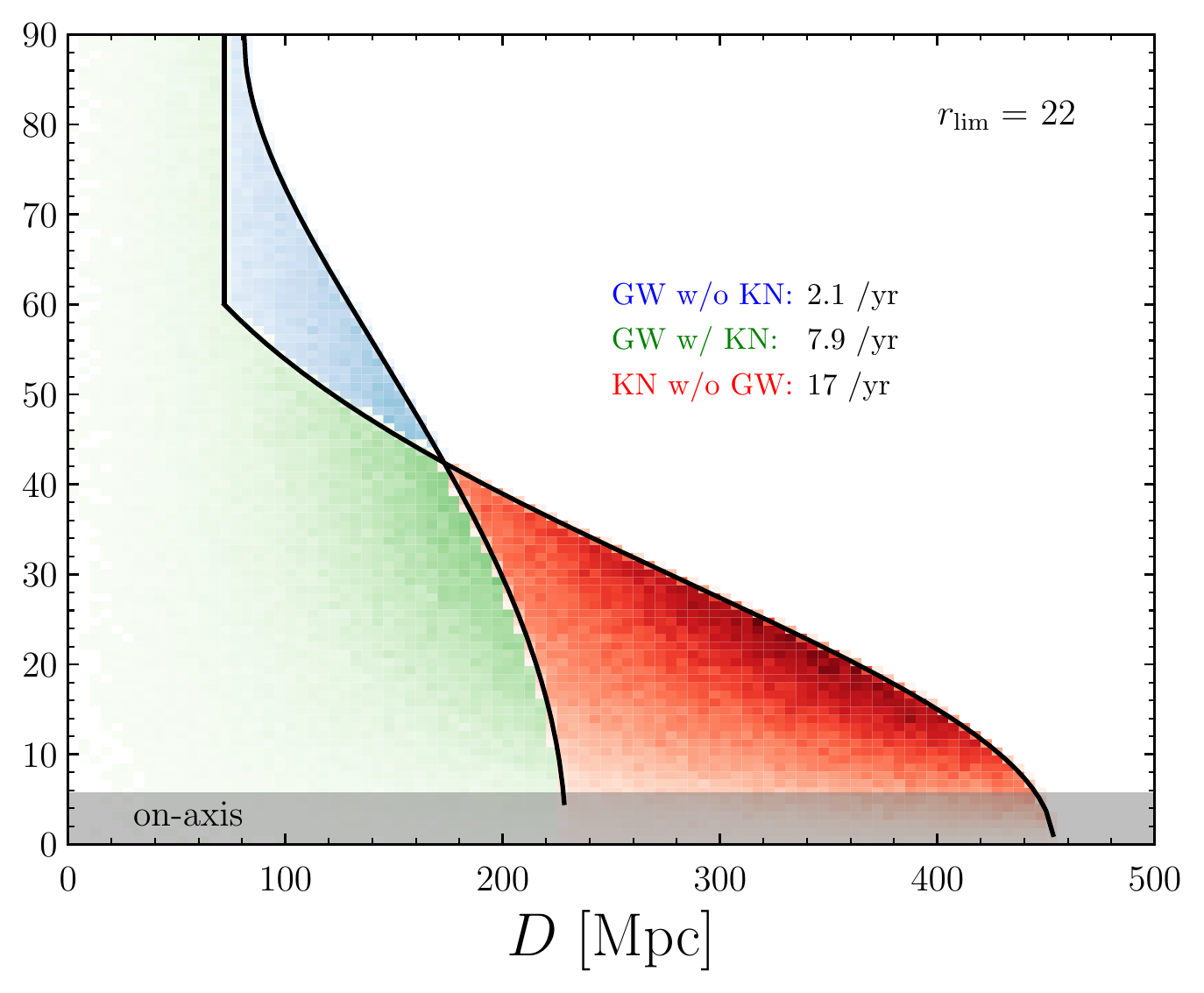}}}
\caption{Relative occurrence rates of signals in the distance-viewing angle plane predicted for the O4 run. Colors indicate different detection scenarios: events detectable (i) both as GW triggers and kilonovae (green), (ii) as GW signals alone (blue), and (iii) only as kilonovae (``orphan kilonovae,'' red). For the kilonova detection, the four diagrams correspond to the limiting $r$-band magnitudes from 19 to 22. We also indicate the total occurrence rates in each detection scenario, assuming a GW detection rate of 10 yr$^{-1}$ for O4.}
\label{fig:GWKNplane}
\end{figure*}

In Fig.~\ref{fig:GWKNplane} we study the distributions in distance and viewing angles for events detected in the gravitational or optical domains. For this figure only, we 
remove the intrinsic variability of kilonovae introduced in Eq.~\ref{eq:KNmag} (i.e., we set $\delta M_\lambda$ to $0$) to clearly separate the different observing scenarios in the distance-viewing angle plane.

For limiting $r$-band magnitudes equal to or smaller than 20, practically all kilonovae that can be detected will follow a GW event if the interferometers are taking data at the corresponding time. Conversely, for deeper searches reaching $r_{\rm lim} =$ 21 or 22, the fraction of ``orphan kilonovae'' without a GW alert increases and becomes dominant. 

Recently, an archival study searching for kilonovae in 23 months of ZTF data was carried out. Down to a limiting magnitude of $r_{\rm lim} \sim 20.5$ for source detection, no transient consistent with being a kilonova was identified \citep{2020ApJ...904..155A}. Considering that a kilonova can be safely detected and characterized only if its peak magnitude is at least one magnitude brighter than the limit of the survey,  Fig.~\ref{fig:GWKNplane} allows us to estimate the number of expected kilonova detections over the 23 month period.
Assuming perfect identification and a sky coverage of $\sim~50\%,$ as appropriate for ZTF, we find
between 0.4 and 2.6 
detections,
taking into account the uncertainty on the binary neutron star merger rate but not that linked to the kilonova model.

Beyond the kilonova model uncertainties, an overestimated rate of mergers or the limitations of the kilonova identification algorithm as discussed in \citealt{2020ApJ...904..155A} could also contribute to the non-detection. Future surveys and archival studies by other optical facilities \citep{Almualla,Setzer} should clarify which of these options is the most likely.

GRB170817A, associated with GW170817, was very weak considering the distance \citep{Fermi} and cannot be considered as the off-axis view of a bright short GRB \citep{MNP2019}. It was not produced by the core ultra-relativistic jet revealed by very long baseline interferometry observations \citep{Mooley_VLBI,Ghirlanda}; it was instead emitted from the material located at a higher latitude that is propagating toward us \citep{MNP2019}, though not necessarily via the same mechanism as that of other short GRB that are observed on-axis at large distances\LEt{ Verify that your intended meaning has not been changed.}. Therefore, the question of the short GRB-merger connection remains open, even if the evidence for the production of an ultra-relativistic jet in GRB170817 is clearly in line with this hypothesis.

The detection of a bright short GRB seen on-axis following GWs from a binary neutron star merger would represent direct evidence for this connection. Figure~\ref{fig:GWKNplane} allows us to discuss the probability of such events in the future. With  a minimum peak luminosity of $\sim\,10^{50}\,\mathrm{erg/s}$ 
\citep{WP2015,G16}
and a peak energy on the order of $1\, \mathrm{MeV}$ \citep{N07}, short GRBs seen on-axis are bright at any distance below 600 Mpc (peak flux on the order of $1\, \mathrm{ph/cm^2/s}$), and the main limitation for detection by gamma-ray satellites is their sky coverage. Assuming a typical jet opening angle $\theta_j=0.1\, \mathrm{rad}$ \citep[see e.g.,][]{FBMZ2015,67}, Fig.~\ref{fig:GWKNplane} clearly indicates that triple associations of GWs with kilonovae and bright short GRBs seen on-axis should remain especially rare: one event every 5--20 years in the whole sky, according to our calculations.

The association of a bright short GRB with a kilonova even without a GW detection is also a solid argument in favor of the merger connection. Figure~\ref{fig:GWKNplane} shows that the rate of such a double association is more optimistic if the limiting magnitude in the $r$ band is at least 21, with $\sim\,2$ such events per year according to our calculations. 
However, for such bright GRB associations, it has been noted that the optical kilonova signal should only outshine the afterglow flux in dense circum-merger media and with less energetic jets, allowing for an early-breaking or dimmer afterglow \citep{2018A&A...620A.131G}. 

GRB130603B and GRB050709 were well within the parameter region allowing for the kilonova to appear \citep{FBMZ2015}, and yet the associated claimed kilonova components \citep{2013Natur.500..547T,2016NatCo...712898J} were only marginally brighter than the afterglow and required a follow-up duration of longer than a week to be detected.
Still, the potential of such sources for studying binary neutron star merger physics and the larger distances to which these can be detected encourages deep photometric follow-ups of short GRBs.

\subsection{Viewing angle-magnitude diagram for GW190425}
\label{sec:4.4}
GW190425, the only confirmed binary neutron star merger during LVKC observing run O3, was located at $159^{+69}_{-71}\,{\rm Mpc}$ \citep{Abbott_KN}. No kilonova was found during the follow-up that was conducted by several facilities. The deepest searches were led by ZTF to mag. 21 in the $g$ and $r$ bands, covering 21\% of the probability enclosed in the very large final GW sky map of nearly $7500\,\degree^2$ \citep{Kas,ZTF}, and by Pan-STARRS to mag. 21.5 in the $i$ band, covering 28\% of the initial GW sky map \citep{PS1, PS2}.

\begin{figure*}
\resizebox{\hsize}{!}{\includegraphics{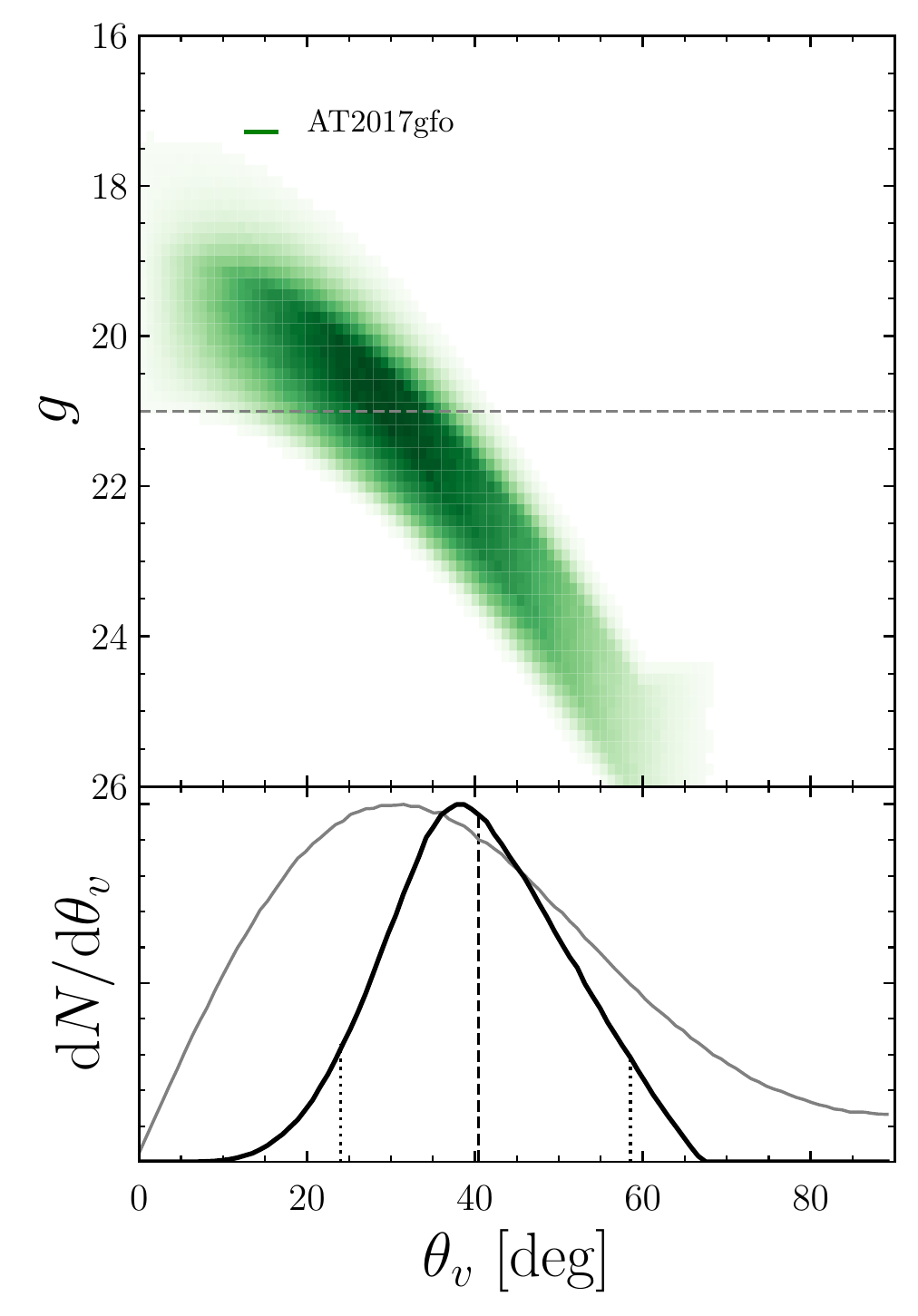}{\includegraphics{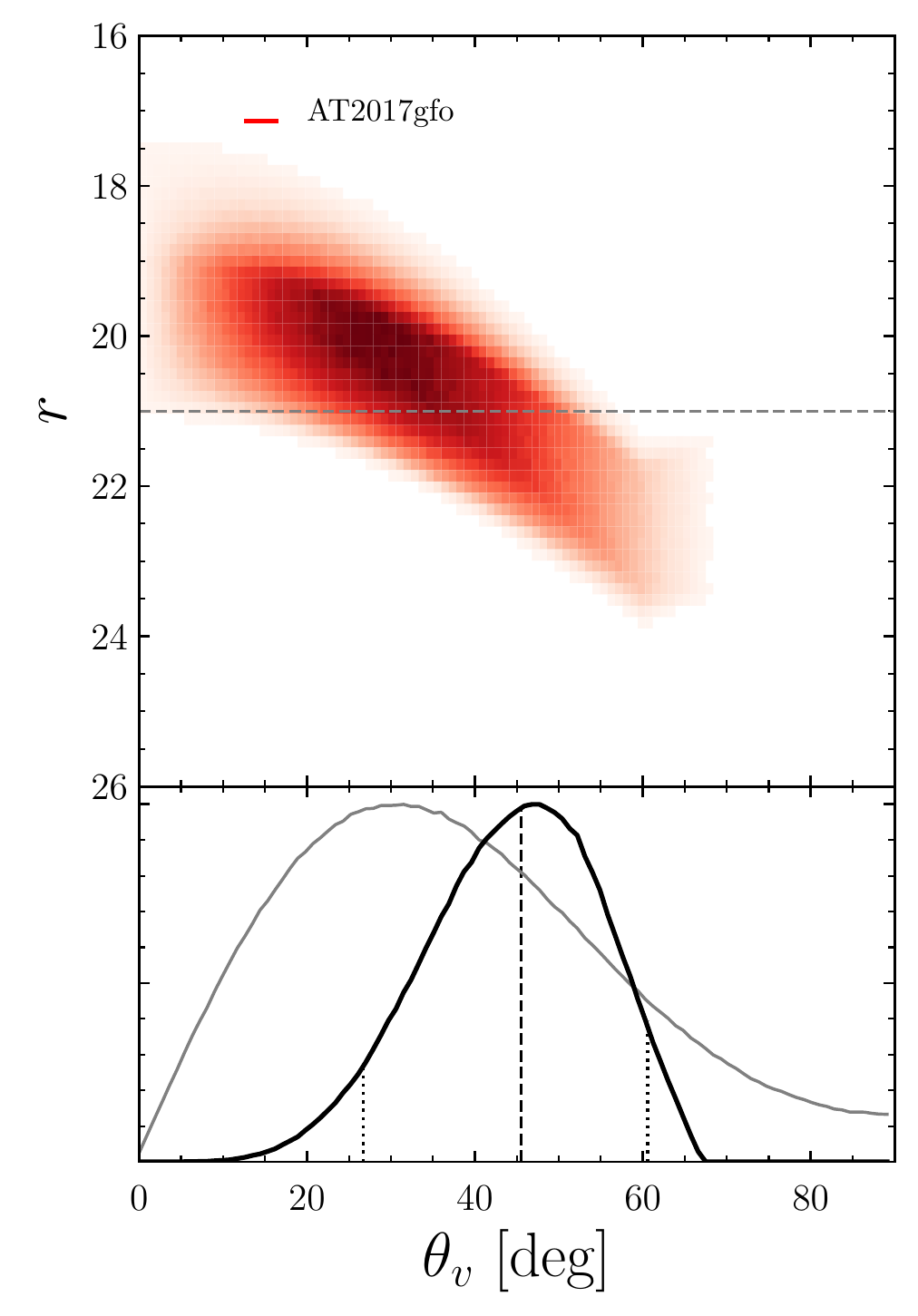}{\includegraphics{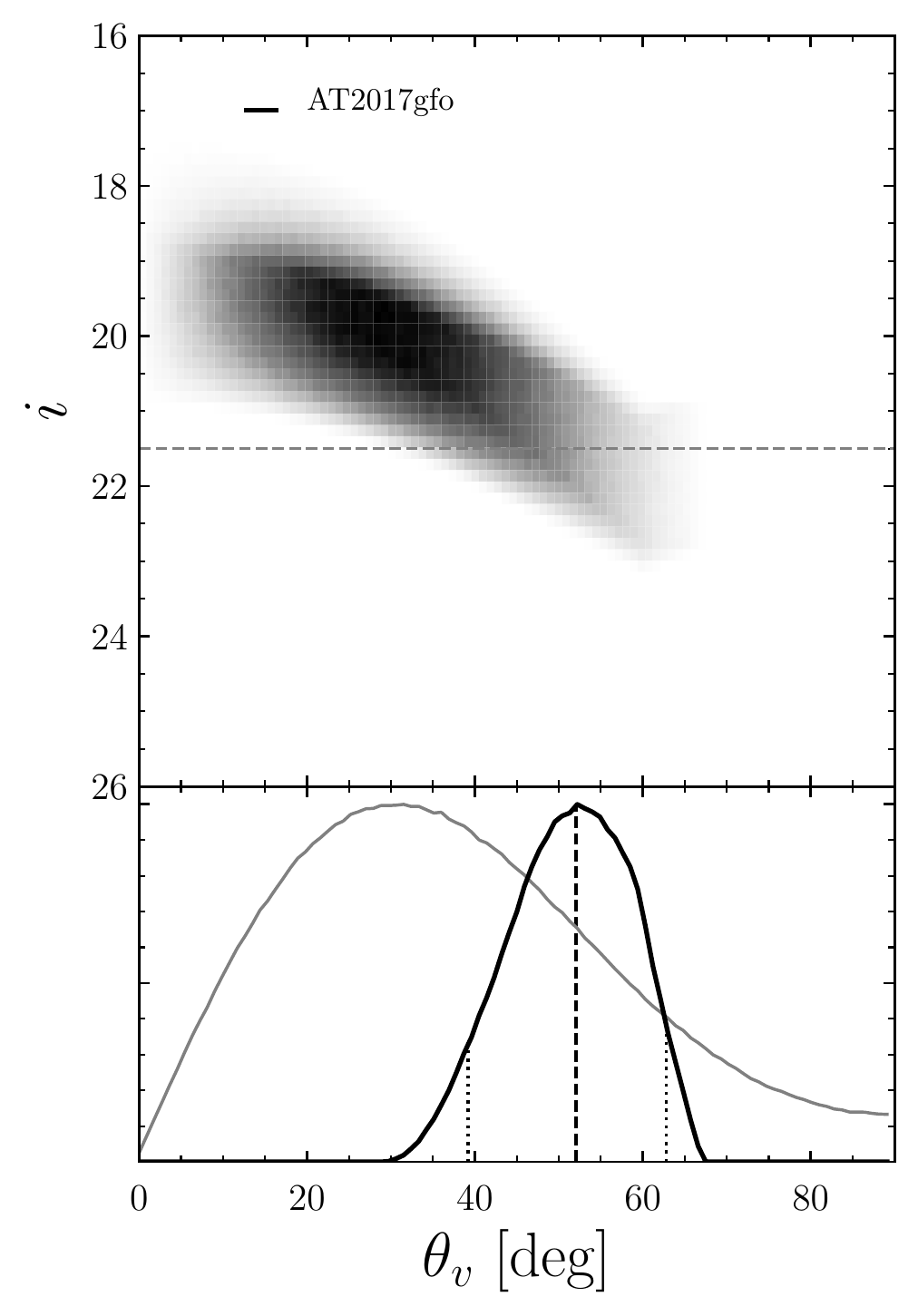}}}}
\caption{Constraints on the viewing angle of GW190425 stemming from the non-detection of its kilonova. Top: Viewing angle--magnitude diagrams in the $g$, $r$, and $i$ bands for events with distances consistent with GW190425 during O3. The horizontal lines are limiting magnitudes for ZTF ($g$ and $r$ bands) and Pan-STARRS ($i$ band). Bottom: Constraints on the viewing angle to GW190425 assuming it was below the  detection limits (black), compared to the viewing angle distribution of all GW triggers (gray).}
\label{fig:GW190425}
\end{figure*}

This non-detection can have different origins, the most obvious being simply that the kilonova was not located in the search area. But it is also interesting to explore the possibility that the kilonova was there but below the detection limit. This possibility is supported by recent modeling predicting that the kilonova of GW190425 was dimmer and faster decaying than AT~2017gfo because of its larger mass \citep{2021arXiv210202229N}. Figure~\ref{fig:GW190425} illustrates the resulting constraints in viewing angle--magnitude diagrams, adopting Eq.~\ref{eq:KNmag} for the kilonova magnitude. For this particularly massive event, we increased the horizon accordingly with the larger chirp mass of GW190425 with respect to a 1.4+1.4 $\Msol$ system, following the relation $D_H \propto \mathcal{M}^{5/6}$ (\citealt{Schutz}; see the value in our Table~\ref{tab:KNparam}).

The viewing angle to GW190425 is bounded below by the non-detection of the kilonova and bounded above by the detection of the GW signal. This results in the individual constraints from the $g$, $r$, and $i$ bands that can be read off Fig.~\ref{fig:GW190425}. The strongest constraint comes from the $i$ band, in which the kilonova is expected to be above typical follow-up thresholds for the largest viewing angle range, according to our model. The combined three-band constraint is $\theta_v^{190425} = 53.3^{+9.8}_{-12.4}\,\degree$, to which a systematic uncertainty due to the kilonova model should be added (see Sect.~\ref{sec:discussion}). Finally, in the case where no blue kilonova was produced in that event -- possibly because the central core of the merged object directly collapsed into a black hole -- no useful constraint can likely be obtained. This last possibility is indeed worth considering because of the large masses of the two neutron stars inferred for GW190425.

\section{Detecting the radio afterglow}
\label{sec:radioaft}
When the kilonova is found, the location of the source is known with an arcsecond accuracy, which allows an efficient search for the afterglow. Assuming an index $p=2.2$ for the power-law distribution of the shock-accelerated electrons, the peak flux of the radio afterglow at 3 GHz is given by \citet{Nakar} and \citet{Gottlieb}:
\begin{equation}
F_{3{\rm GHz}} \sim 8.6\,\varphi\,D_{100}^{-2}\ {\rm max}(\theta_{v, -1},\theta_{j,-1})^{-4.4}\,{\rm mJy} 
\label{eq:AGflux}
,\end{equation}
where $\varphi =E_{52}\theta_{j,-1}^{2}n_{-3}^{0.8}\,\epsilon_{e,-1}^{1.2}\,\epsilon_{B,-3}^{0.8}$ collects the flux dependences on the parameters not related to the observing conditions. Here $E_{52}$ and $\theta_{j,-1}$ are the isotropic energy and opening angle of the jet core (in units of $10^{52}$ erg and 0.1 rad), $n_{-3}$ is the density of the external medium (in units of $10^{-3}$
cm$^{-3}$), $\epsilon_{e,-1}$ and $\epsilon_{B,-3}$ are the usual microphysics parameters (in units of 0.1 and $10^{-3}$), and $D_{100}$ is the distance of the source (in units of 100\,Mpc). The normalization of $\varphi$ was chosen such that $\varphi = 1$ for GW170817-like afterglows \citep[e.g.,][]{2017MNRAS.472.4953L,RSIMB+2018,PhysRevLett.120.241103,TVRRB+2019}.

We find that for O4 and with $\varphi=1$, the afterglow can be detected at the VLA 3 GHz sensitivity of 15 $\mu$Jy for 37\%, 56\%, and 76\% of kilonovae, assuming $r$-band limiting magnitudes of 21, 20, and 19, respectively, for the kilonova search. To construct an exploitable light curve, we can impose a radio flux threshold of three times the detection limit. In this case, these fractions become 23\%, 36\%, and 53\%. In terms of absolute numbers, they respectively correspond to 1.1, 0.7, and 0.3 joint GW-kilonova-afterglow detections per year.

For particularly energetic jets or dense circum-merger environments (i.e., with $\varphi = 10$), the fractions of kilonovae with radio afterglows at three times the VLA limit rise to 59\%, 81\%, and 97\%. This corresponds to 2.9, 1.6, and 0.5 three-signal detections per year.
\LEt{ single-sentence paragraph.}

\begin{figure*}
\centering
\resizebox{\hsize}{!}{\includegraphics{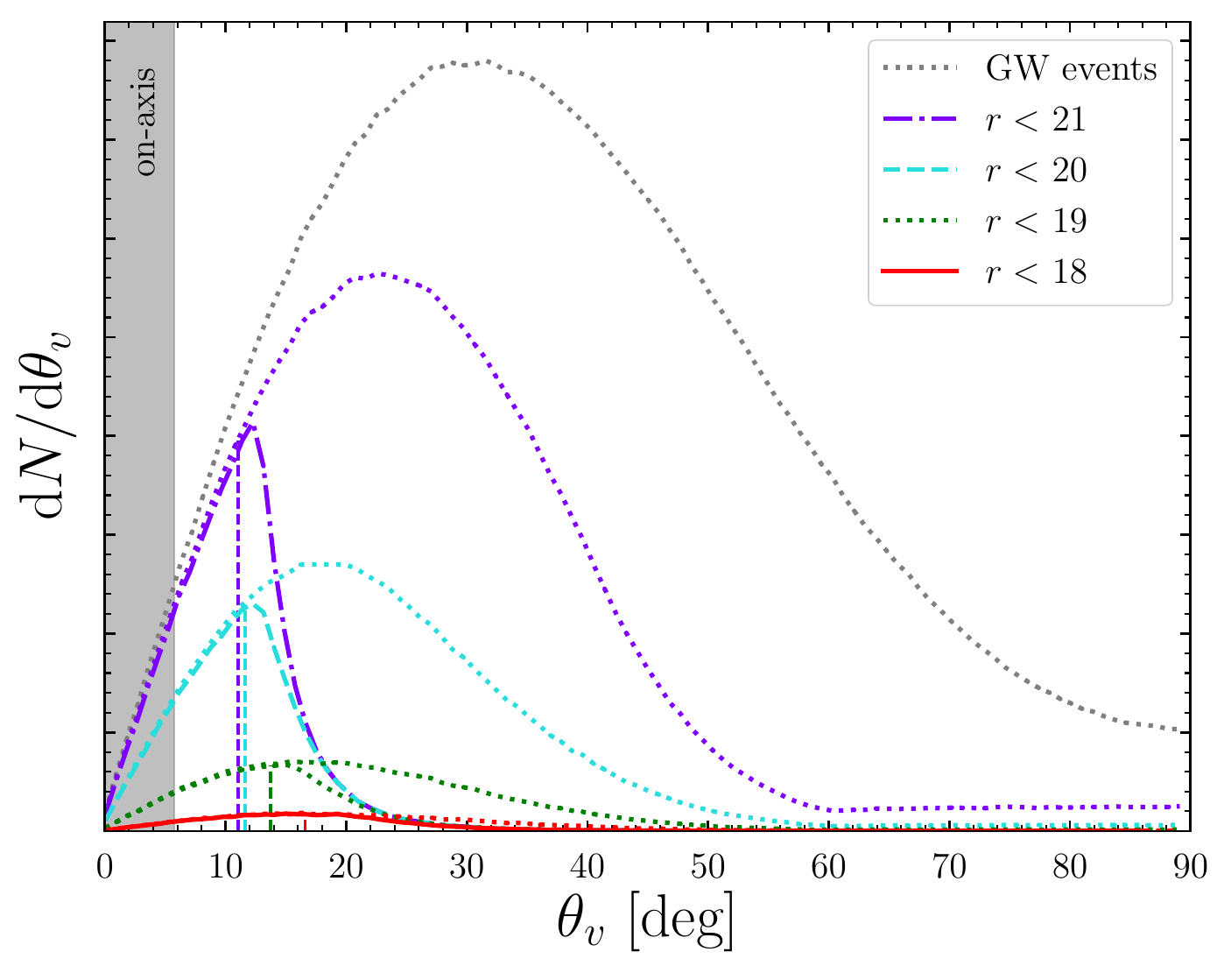}{\includegraphics{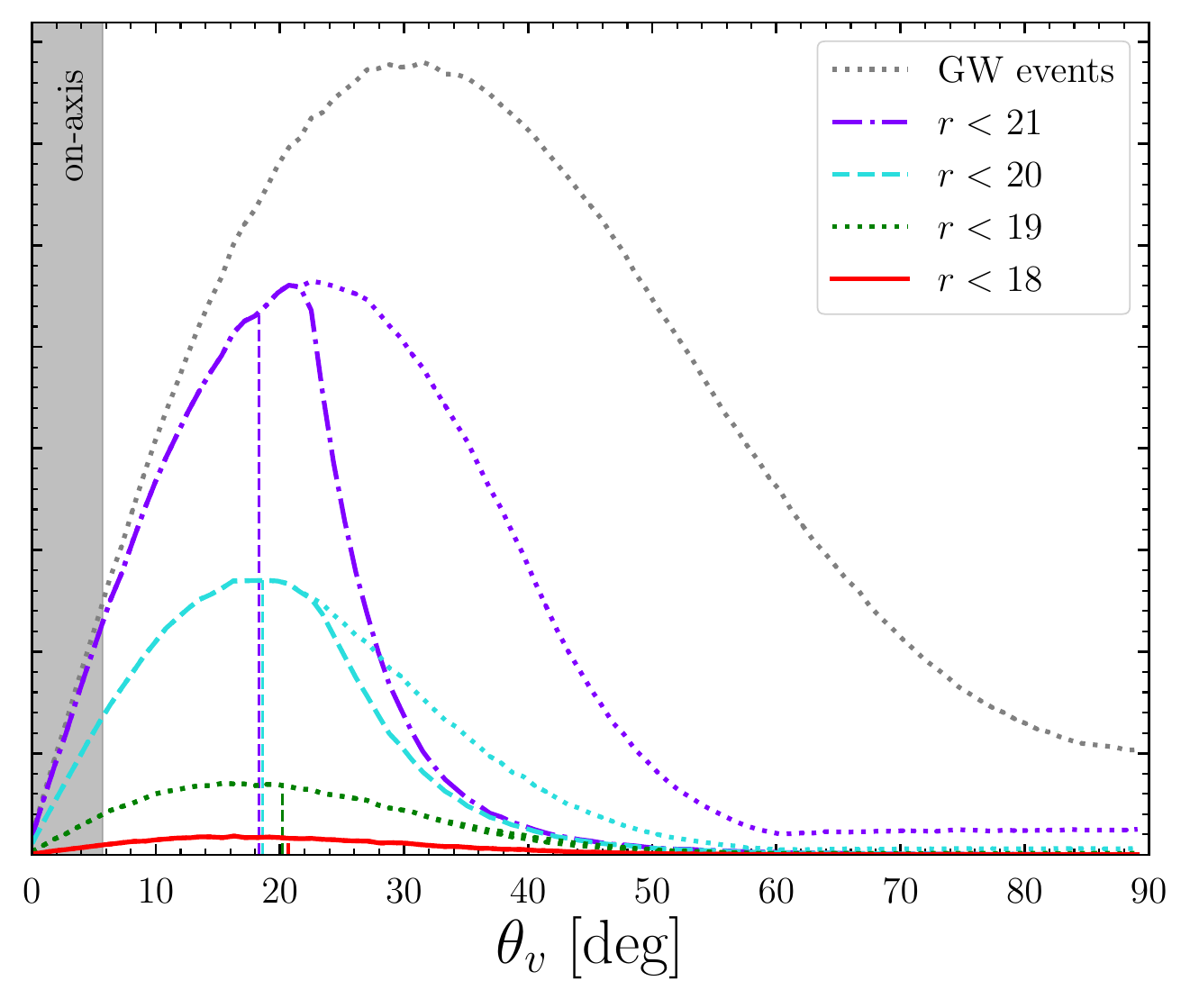}}}
\caption{Distributions in viewing angle of the afterglows detectable in the radio
band at three times the VLA threshold (45 $\mu$Jy) following a GW-triggered
kilonova detection with a limiting $r$-band magnitude of 18 to 21. The dotted
lines represent the corresponding distributions for the kilonova sources, as in
Fig.~\ref{fig:KNorientation}. In the left (resp. right) panel, a value $\varphi=1$ (resp. $\varphi=10$, corresponding to particularly energetic jets or dense circum-merger media) was adopted in Eq.~\ref{eq:AGflux}.}
\label{fig:AG}
\end{figure*}

We represent in Fig.~\ref{fig:AG} the distribution in viewing angle of the afterglows that can be detected with the VLA at three times the detection limit for $\varphi = 1$ and 10 and different kilonova search limiting magnitudes. Due to the very strong dependence of the afterglow flux on the viewing angle ($F\propto \theta_v^{-4.4}$ for $p=2.2$; see Eq.~\ref{eq:AGflux}), the detection is only possible at small viewing angles, $<15-20\,\degree$.

\section{Discussion}
\label{sec:discussion}

We have presented a population model for kilonova counterparts to GW binary neutron star merger signals. In Sect.~\ref{sec:2} we obtained the distributions in distance and viewing angle of the GW-trigger events (see Fig.~\ref{fig:GW}). Then, using a simple parametrization of the kilonova peak AB magnitude in various spectral bands as a function of viewing angle, we computed the distributions of sources in three magnitude intervals (<18, 18--20, and >20) for O3 and the future O4 and O5 observing runs. We also considered the possibility that some kilonovae lack the blue component due to the lack of some mass ejection episodes during the merger.

The rate of kilonovae brighter than a given limiting magnitude was obtained (see Eq.~\ref{eq:KNrate} and Fig.~\ref{fig:KNmag}b). This confirms the extraordinary chance of having observed the August 17 event so early. In Fig.~\ref{fig:KNorientation} we studied the viewing angles of detectable kilonovae for different limiting magnitudes. The median of this distribution, about $36\,\degree$ for the GW trigger population, decreases as the search becomes shallower, reaching $26\,\degree$ and then $21\,\degree$ for $r$-band limiting magnitudes of 21 and lower than 20.

We then studied the regions in distance--viewing angle space where the gravitational signal or the kilonova can be detected (see Fig.~\ref{fig:GWKNplane}). In each zone, we estimated the event rate, normalized to a GW trigger rate of $10\,{\rm yr}^{-1}$, as expected for O4. For deep surveys reaching mag. 21--22, the rate of orphan kilonovae \LEt{ Quotation marks can be used to introduce a special meaning for a word or phrase the first time it is used in the text, but thereafter quotation marks are not needed.}without a detectable GW signal becomes dominant, opening the way to detecting kilonova counterparts to short GRBs, which should become much more common. The progress in understanding the merger phenomenon by leveraging such signals motivates the effort to carry out these surveys and the optical follow-up of short GRBs.

As the sensitivity of GW detectors increases, more distant events will be found, but the rate of bright kilonovae ($r < 19$) will not change. Conversely, follow-ups reaching magnitudes 20--21 have the potential to find five to ten times more kilonovae beyond magnitude 19, leading to a potential discovery rate of ten events per year during O4 to more than twenty for O5 (see Fig.~\ref{fig:KNmag}b). Obviously, going from potential to effective discoveries will require a deep inspection of the GW sky map by target-of-opportunity endeavors or the analysis of untargeted searches by high cadence facilities, such as the Vera C. Rubin facility \citep{Margutti,LSST}. 

We studied the case of radio afterglow counterparts to kilonova signals. We note that our results in this respect are consistent with our prior study of afterglow counterparts \citep{Duque}, during which the kilonova was not considered. That study and the present paper find consistent results in the limit of very deep kilonova search limiting magnitudes. However, the present paper shows that, as of O4 and with current follow-up capacities, the kilonova detection will become a limiting factor in the search for afterglow counterparts.

We also used our kilonova model to constrain the viewing angle of GW190425, the only confirmed binary neutron star merger event since GW170817. No associated kilonova was detected; this could simply be a consequence of the poor localization, which limited the search to less than 30\% of the GW sky map. If, however, a kilonova was in the search area but was too weak to be detected, a constraint on the viewing angle can be obtained.\ We find that the viewing angle must have been $53\pm 10\, \degree$, assuming there was an AT~2017gfo-like blue component.

In our kilonova model there are  two sources of uncertainty: one linked to the polar-to-equatorial view contrast ($\Delta M_\lambda$ in Eq.~\ref{eq:KNmag}) and one linked to the calibration of the polar magnitude ($M_\lambda$)\LEt{ Verify that your intended meaning has not been changed.}. The former was fit to \LEt{ to?}theoretical expectations from sophisticated modeling and the latter from calibration on GW170817. All of our results are sensitive to these procedures, though we underline that our general conclusions (see Sect.~\ref{sec:conclusion}) should remain valid and our constraints on GW190425 should be seen as proof of concept. This method will be most useful in the case of genuine non-detections when the future, smaller GW sky maps will effectively be fully covered by follow-up. For future GW triggers, one could use the GW-measured progenitor properties (such as component masses and tidal deformability) to tailor the kilonova modeling to the specific events \citep{2021arXiv210202229N}. In case of a non-detection, the viewing angle constraints thus obtained would be more robust because they are informed with the complete multi-messenger data set.

Both aspects of this uncertainty should improve in the coming years with the detection and observation of even a limited sample of kilonovae following GW signals, allowing us to explore both their intrinsic diversity and their properties under different viewing angles \LEt{ or "allowing both their intrinsic diversity and their properties to be explored under different viewing angles.}. When the burst afterglow is also detected, information on the external density and a better estimate of the viewing angle can be obtained, which might be completed, in the long term, with the potential observation and leveraging of the kilonova afterglow \citep{82, 83, 188}. 

Finally, as we have noted, kilonovae and even mild associated viewing angle measurements seem to be the only means for electromagnetic modeling to contribute to multi-messenger cosmology and the resolution of the Hubble tension \citep{2020arXiv201212836M}. The other counterparts are ruled out for their rareness. The effort to collect a kilonova sample and study source variability and viewing angle properties thus appears even more desirable in this regard.

\section{Conclusion}
\label{sec:conclusion}

We have presented a population study of kilonova counterparts to GW binary neutron star merger signals based on a simple viewing-angle-dependent model deduced from state-of-the-art modeling and calibrated on AT~2017gfo.

For shallow searches, the rate and viewing angle properties of the detected kilonova population are independent of the GW sensitivity. However, deep searches by target-of-opportunity endeavors and high-cadence surveys can probe a high-inclination population, detecting tens of events per year with design-type GW observing runs. Deep surveys will, however, be dominated by non-GW-triggered (orphan) kilonovae with possible short GRB associations.

We have proven the concept of constraining the inclination angle of systems in the case of a non-detection of the kilonova counterpart. Our method will become more effective in the case of a genuine non-detection when future, smaller GW sky maps are fully covered by follow-up.

Our results would be refined with a better understanding of kilonova emissions, and dedicated efforts to collect a sample are warranted. Such efforts are further motivated given the potential role of kilonovae in precision multi-messenger cosmology.

\section*{Acknowledgments}
We thank G. Duque for providing software hosting and running resources to this project.

We acknowledge the Centre National d'Études Spatiales (CNES) for financial support in this research project. Finally, we thank the referee for a constructive report which improved the quality of the manuscript.

\bibliographystyle{aa}
\bibliography{kn_paper}
\end{document}